\documentclass[structabstract]{aa}  
\usepackage{graphicx}
\usepackage{txfonts}
\newcommand{\eg}{{\it e.g.}}
\newcommand{\ie}{{\it i.e.}}
\newcommand{\viz}{{\it viz.}}
\newcommand{\etal}{{\it et al.}}

\newcommand{\Msun}{\ensuremath{M_{\odot}}}
\begin{document}
   \title{Young pre-Low-Mass X-ray Binaries in the propeller phase}

   \subtitle{Nature of the 6.7-hour periodic X-ray source 1E 161348-5055 in RCW 103}

   \author{Harshal Bhadkamkar
          \and
          Pranab Ghosh
          }

   \institute{Department of Astronomy \& Astrophysics, Tata Institute
     of Fundamental Research, Mumbai 400 005, India\\
              \email{pranab@tifr.res.in, harshalhb@tifr.res.in}     
             }

   \date{Received; accepted}

 
  \abstract
   {Discovery of the 6.7-hour periodicity in the X-ray source 1E
     161348-5055 in RCW 103 has led to investigations of the nature of
    this periodicity.}
   {To explore a model for 1E 161348-5055 wherein a fast-spinning
     neutron star with a magnetic field $\sim 10^{12}$ G in a young 
     pre-Low-Mass X-ray Binary (pre-LMXB) with an eccentric orbit 
     of period 6.7 hr operates in the ``propeller'' phase.}
   {The 6.7-hr light curve of 1E 161348-5055 is modeled in terms of
     orbitally-modulated mass transfer through a viscous accretion
     disk and subsequent propeller emission. Formation of eccentric 
     binaries in supernovae and their subsequent tidal evolution are
     studied.}
   {The light curve of 1E 161348-5055 can be quantitatively accounted
   for by models of propeller torques of both Illarionov-Sunyaev 
   type and Romanova-Lovelace \etal~type, and spectral and other 
   properties are also in agreement. Formation and evolution of model 
   systems are shown to be in accordance both with standard theories 
   and with X-ray observations of 1E 161348-5055.}
   {The pre-LMXB model for 1E 161348-5055 and similar sources agrees
   with observation. Distinguishing features between this model and 
   the recently-proposed magnetar model need to be explored.}


   \maketitle

\section{Introduction}
\label{sec:intro}

The point soft X-ray source 1E 161348-5055 (henceforth 1E) near the 
center of the young ($\sim 2000$ yr old) supernova remnant (SNR) 
RCW 103 has attracted much attention lately, following the discovery 
of a strong 6.67 hr periodic modulation in 1E by de Luca \etal~(2006,
henceforth dL06) from a deep \emph{XMM-Newton} 
observation of the source in 2005. 1E was 
discovered in 1980 (\cite{garmire}) as a soft \emph{Einstein} X-ray source. The 
original interpretation as an isolated neutron star was found to
be untenable in view of subsequent discovery by other X-ray satellites
(\eg, \emph{ROSAT, ASCA, Chandra}) 
of the large variability of 1E on the timesacle of a few years (dL06).
A periodicity at $\sim 6$ hr was first hinted at by \emph{Chandra}
observations, but the first clear, strong detection came from the 
above 2005 observations of
dL06, who also showed the existence of this periodicity in the data
from earlier 2001 observations of 1E with \emph{XMM-Newton}, when
the source luminosity was higher by a factor $\sim 6$ during the
course of its sequence of several-year timescale outbursts nentioned
above, documented by these authors from archival data.

The nature of the above 6.67 hr periodicity is an interesting
question, on which preliminary discussions were reported in dL06.
Recently, Pizzolato \etal~(2008, henceforth P08) have proposed a model
for the 1E system wherein it is a close binary consisting of a magnetar, 
\ie, a neutron star with a superstrong magnetic field $\sim 10^{15}$
G, and a low-mass companion. The 6.67 hr periodicity is identified in 
this model with the spin period of the neutron star, 
to which this young neutron star has been spun down in such a
short time by the torques associated with its enormous magnetic field.
This period has also been proposed by P08 to be in close synchronism
with with the orbital period of the binary, in analogy with what is
believed to be happening in Polar Cataclysmic Variables or AM Her-type
systems. The observed X-ray emission from 1E is that from the magnetar
in this model.        

In this paper, we explore an alternative model for the 1E system
wherein it is a close binary system consisting of a young neutron star with
a canonical magnetic field $\sim 10^{12}$ G, and a low-mass companion,
\ie, a pre-Low-mass X-ray Binary (henceforth pre-LMXB), such as are 
believed to be the standard progenitors of Low-mass X-ray Binaries 
(henceforth LMXBs). Such pre-LMXBs are born after 
the common-envelope (CE) evolution phase of the original progenitor
binary system consisting of a massive star and a low-mass companion,
which leads to the formation of a binary consisting of the He-core
of the original massive star and the low-mass companion
(\cite{ghoshbook} and references therein). 
The He-star susequently explodes in a supernova, leading to a 
neutron star in orbit with a low-mass companion, \ie, the pre-LMXB
referred to above. This is the standard He-star supernova scenario
for the formation of LMXBs (\cite{ghoshbook} and references therein). 
The 6.67 hr periodicity is identified in our model 
with the orbital period of the binary. In our model the young neutron 
star is still spinning very rapidly, with a canonical spin period 
$\sim 10-100$ ms, and is operating in the ``propeller'' regime, 
wherein any matter approaching the fast-rotating magnetosphere of
the neutron star is expelled by the energy and angular momentum 
deposited into it through its interaction with the magnetospheric
boundary (\cite{is}, henceforth IS75, \cite{dav1, dav2}, \cite{ik}, 
\cite{mrf}, \cite{iik}, \cite{ghoshprop} and references therein,
 henceforth G95, \cite{love}, henceforth LRB99 \cite{rom04}, 
\cite{rom05}, henceforth RUKL05, \cite{usty}, henceforth UKRL06). 

The observed X-ray emission from 1E in our model is that from the 
propeller: indeed, it is well-known that soft X-ray transients 
(SXRTs) like Aquila X-1 and others (see Sec.~\ref{sec:spectra}) go through
low/quiescent states during the decay of their outbursts, during
which their luminosities and spectral properties are very
similar to those of 1E, and the neutron stars in them are 
believed to be in the propeller regime (\cite{campana}, \cite{stella}). 
The observed 6.67 hr periodicity in our model is due to the 
orbital modulation of the supersonic propeller, which is 
caused by the orbital modulation of the mass-transfer rate in
the \emph{eccentric} binary orbit of a young system like 1E.
It is well-known that young post-SN binaries with low-mass
companions like 1E are almost certain to have eccentric orbits,
due to the large eccentricities produced in such systems in
the SN explosion (see Sec.~\ref{sec:postsn}) and the duration of 
the subsequent tidal circularization compared to the ages of
systems like 1E (see Sec. \ref{sec:tidal}). By contrast, SXRTs are
believed to be old LMXB systems with circular orbits, where
such modulation will not occur. 

We show in this work that the 6.67 hr light curve of 1E can be 
accounted for quantitatively by our model for propeller torques
of both Illarionov-Sunyaev type and Romanova-Lovelace \etal~type 
(see Sec.~\ref{sec:propeller}), and that the observed 
spectral and other characteristics are also in general agreement 
with our overall picture. Thus, further diagnostic features need
to be explored in order to distinguish between our model and the
magnetar model as a viable description of this and similar sources.
           
\section{Propeller phase in pre-LMXBs}
\label{sec:propeller}
   
In a pre-Low-Mass X-ray Binary (pre-LMXB: see above), 
the newborn, fast-rotating neutron star     
is unable at first to accrete the matter that is being transferred
from the companion through the inner Lagrangian point $L_1$, 
because of the fast rotation of the neutron star (IS75, 
\cite{dav1, dav2}, \cite{ik}, \cite{mrf}, \cite{iik}, G95, LRB99,
\cite{rom04}, RUKL05, UKRL06). 
Because of its large angular momentum, this matter forms an accretion 
disk and reaches the magnetospheric boundary of the magnetized neutron 
star, whereupon this ionized matter interacts with the fast-rotating 
neutron star's magnetic field, and the energy and angular momentum 
deposited in it by magnetic stresses associated with this 
fast-rotating magnetic field expel it. This is the 
\emph{propeller phase} of the system (IS75), during which the 
neutron star spins down as it loses angular momentum and rotational  
energy. During this propeller phase, the disk matter at the magnetospheric 
boundary is shock-heated as the ``vanes'' of the supersonic propeller
(IS75) hit it, and the hot matter emits in the soft
X-ray band. This emission appears unmodulated at the neutron-star spin
frequency (as opposed to the X-ray emission from canonical 
accretion-powered pulsars, which comes from the neutron-star surface) 
to a distant observer, who sees only the total emission 
from the heated matter at the magnetospheric boundary. Observations of 
transient low-mass X-ray binaries (\ie, the soft X-ray transients or
SXRTs) like Aquila X-1 (\cite{campana}) and SXJ1808.4-3658
(\cite{stella}) in quiescence, when the neutron stars in them are
thought to be operating in the propeller phase, amply confirm this point.    

The propeller luminosity $L$ during the above phase is given by 
$L=N\omega$, where $N$ is the propeller torque acting on the
neutron star and $\omega$ is its spin angular velocity. 
The propeller torque $N$ was first estimated by IS75 in their pioneering 
suggestion of this mechanism, and subsequent work over approximately 
the next two decades considered variations of this torque under different 
circumstances, as summarized in G95. These works addressed themselves
largely to quasi-spherical accretion, and we shall call this kind of
propeller torque the Illarionov-Sunyaev type (or IS-type for short)
torque, which was widely used in that time-frame in propeller spindown 
calculations. In the 2000s, Romanova, Lovelace and co-authors reported 
a series of calculations of the propeller effect for disk-accreting 
magnetic stars based on their numerical MHD simulations (\cite{rom04}, 
RUKL05, UKRL06; also see the analytic estimates in LRB99). We shall
call the propeller torque obtained from this line of work the 
Romanova-Lovelace \etal~type (or RUKL-type for short) torque. In this
work, we shall consider both IS-type and RUKL-type propeller torques
for the problem at hand.

Consider IS-type torques first. 
For such fast-rotating neutron stars as we are concerned with in this 
work, the propeller operates in the \emph{supersonic} regime, and its 
torque is given by (G95 and the references therein),
\begin{equation}
N = \frac{1}{6}\frac{\mu^{2}\omega^{2}}{GM_x}\frac{\Omega(r_m)}{\omega}.
\label{eq:sstorque}
\end{equation}
In Eq.~(\ref{eq:sstorque}), $\Omega(r_m)$ is the Keplerian angular 
velocity at the magnetospheric radius $r_m$, $\mu$ is the magnetic
moment of neutron star and $M_x$ is its mass. Combining this equation with 
the standard expression for the magnetospheric radius 
(\cite{ghoshbook}), \viz, 
\begin{equation}
r_m = \left[\frac{\mu^{2}}{\dot{M}\sqrt{2GM_x}}\right]^{\frac{2}{7}},
\label{eq:alfvenrad}
\end{equation}
where $\dot{M}$ is the rate at which transferred matter arrives at the
magnetospheric boundary, we obtain the following expression for the 
propeller luminosity:
\begin{equation}
L_{35} \approx 5 \dot{M}_{14}^{\frac{3}{7}}\:(P_{spin}/0.1 
{\rm s})^{-2}\:\mu_{30}^{\frac{8}{7}}\: m_{x}^{\frac{-2}{7}} .
\label{eq:mdotprop}
\end{equation}
In Eq.~(\ref{eq:mdotprop}), $\dot{M}_{14}$ is $\dot{M}$ in units of 
$10^{14}$ g s$^{-1}$, $L_{35}$ is $L$ in the units of $10^{35}$ erg s$^{-1}$,
$P_{spin}$ is the neutron-star spin period, $\mu_{30}$ is the 
neutron-star magnetic moment in units of $10^{30}$ G cm$^3$, and
$m_{x}$ is the neutron-star mass in units of solar mass. As neutron 
stars are thought to have $P_{spin}\sim 0.01-0.1$ s at birth, and 
as the propeller phase is thought to end when the spin period is
longer than  $P_{spin}\sim 0.1-1$ s, we have made the canonical choice
for the expected scale of $P_{spin}$ in systems like 1E. Equation 
(\ref{eq:mdotprop}) clearly shows how the propeller luminosity scales 
with the mass-arrival rate $\dot{M}$, and essential neutron-star 
properties, namely, its spin period $P_{spin}$, its magnetic moment 
$\mu$, and its mass $M_x$.

Now consider RUKL-type torques. These authors summarized the 
results of some of their extensive MHD simulations in RUKL05 and 
UKRL06 in terms of power-law fits to these results, showing that the
scaling of the \emph{total} propeller torque $N$ with the magnetic 
moment $\mu$ and the spin rate $\omega$ of the neutron star was 
\begin{equation}
N \propto \mu^{1.1}\omega^2.
\label{eq:ruklscale1}
\end{equation}
However, the scaling of $N$ with $\dot{M}$ was not available from 
the above references, because only the parameters $\mu$ and $\omega$
(and also the turbulence and magnetic diffusivity parameters of the 
disk: see below) seem to have been varied in the series of 
simulations reported in these references. In order to estimate the
scaling of $N$ with $\dot{M}$ for RUKL-type torques, we proceeded 
in the following way. 

First, we did an analytic estimate in the following manner. In 
their analytic study, LRB99 argued that the radius $r_d$ of the 
inner edge of the disk should depend on the stellar rotation rate
$\omega$ in addition to the parameters $\mu$ and $\dot{M}$ that 
$r_m$ (see above) depended upon. The scaling with $\omega$, $\mu$, 
and $\dot{M}$ that these authors derived was revised in UKRL06, 
the final result being given as $r_d \propto \mu^{1/2}\dot{M}^{-1/4}
\omega^{-1/4}$. (Note the closeness of the scalings with $\mu$ and 
$\dot{M}$ with those which apply to $r_m$, as given above.) 

In a simple first approach, if we argue that a reasonable estimate
of the torque scalings may be obtained by replacing $r_m$ with 
$r_d$ in Eq.~(\ref{eq:sstorque}) for disk accretion, we arrive at
the scaling
\begin{equation}
N \propto \mu^{5/4}\omega^{11/8}\dot{M}^{3/8}
\label{eq:ruklscale2}
\end{equation}        
for RUKL-type torques. Noticing the qualitative similarity of the 
the scalings with $\mu$ and $\omega$ in Eq.~(\ref{eq:ruklscale2}) 
with those of the actual RUKL-type torque given in 
Eq.~(\ref{eq:ruklscale1}), and furthermore the quantitative 
closeness for the scaling with $\mu$, we argued that the best 
estimate would be to use the scalings of Eq.~(\ref{eq:ruklscale1})
for $\mu$ and $\omega$, and the scaling of Eq.~(\ref{eq:ruklscale2})
for $\dot{M}$, thus arriving at a suggested scaling for the 
RUKL-type torque as
\begin{equation}
N \propto \mu^{1.1}\omega^2\dot{M}^{3/8}.
\label{eq:ruklscale3}
\end{equation}    

Before proceeding further, we recognized that RUKL-type torques
may arise from more complicated interactions than are describable 
by the above arguments, and so attempted to verify the above 
$\dot{M}$ scaling by further comparison with RUKL results. To this
end, we noted the correlated variations of $N$ and $\dot{M}$
recorded in Figure 4 of \cite{rom04}, and fitted the two prominent 
peaks in $N$ and $\dot{M}$ at the extreme right of this figure to
a power law. This gave an exponent $\approx 0.37$, coincident with
that in Eq.~(\ref{eq:ruklscale3}) within errors of determination.
With this support, we use the scalings of Eq.~(\ref{eq:ruklscale3}) 
for the RUKL-type torque in this work, deferring further 
considerations to future publications. 

In order to obtain the dimensional values of the RUKL-type propeller 
torques and related varaiables, we now insert the reference units
for the RUKL simulations given in RUKL05 and UKRL06, thus obtaining
for the torque:
\begin{equation}
N_{33} \approx 0.87\mu_{30}^{1.1}(P_{spin}/0.01 {\rm s})^{-2}
\dot{M}_{14}^{3/8}.
\label{eq:rukltorq}
\end{equation}     
Here, $N_{33}$ is the propeller torque in units of $10^{33}$ g cm$^2$
s$^{-2}$, the units of other variables are as before, and we have 
kept the values of the turbulence and magnetic diffusivity parameters 
of the accretion disk in RUKL-type models at the canonical values
given in RUKL05 and UKRL06. The RUKL propeller luminosity $L$ is 
then obtained in a straightforward manner as
\begin{equation}
L_{35} \approx 5.5 \dot{M}_{14}^{\frac{3}{8}}\:(P_{spin}/0.01 
{\rm s})^{-3}\:\mu_{30}^{1.1}.
\label{eq:luminrukl}
\end{equation}
In Eq.~(\ref{eq:luminrukl}), the units of all variables are as 
before.

Comparison of Eqs.~(\ref{eq:mdotprop}) and (\ref{eq:luminrukl})
immediately leads to the following conclusions about IS-type and
RUKL-type propeller luminosities. First, the scalings with $\mu$ 
and $\dot{M}$ are almost identical for the two types. Secondly,
the scaling with the neutron-star spin period $P_{spin}$ is stronger
(-3 instead of -2) for the RUKL-type than for the IS-type. Finally,
for identical values of the variables $\mu$, $\dot{M}$, and  
$P_{spin}$, the RUKL-type propeller luminosity is about three orders
of magnitude lower than the IS-type propeller luminosity. 
Conversely, at fixed values of $\mu$ and $\dot{M}$, roughly equal
luminosities are given by the two types if the spin-rate for the
RUKL type is about an order of magnitude higher than that for the
IS type. 
  
As indicated earlier, in this work we are exploring the properties of 
such propellers as described above during the relatively early 
stages of post-supernova binaries containing pre-LMXBs, when the
binary orbits are expected to be appreciably eccentric, as explained
in Sec.~\ref{sec:postsn}. In such a system, the mass-transfer rate 
$\dot{M}_{tr}$ through the inner Lagrangian point $L_1$ is expected to 
vary periodically with the orbital phase, as detailed below in 
Sec.~\ref{sec:masstransfer}.
This flow of matter forms an accretion disk because of its large
specific angular momentum, as explained above, and slow viscous effects
in the disk modify the profile of the above periodic modulation
(making it less sharp), and the resultant periodic profile is that
which is shown by the mass-arrival rate $\dot{M}$ at the neutron
star. The propeller luminosity then follows suite, showing a periodic 
modulation, as described by Eq.~(\ref{eq:mdotprop}) for the 
IS-type torque or Eq.~(\ref{eq:luminrukl}) for the RUKL-type torque.
In this scenario, therefore, we identify the 6.67 hour period of 1E 
with the binary period of a young, eccentric pre-LMXB, which is
expected to turn much later into a standard LMXB after passing through 
further intermediate phases (see Sec.~\ref{sec:recontact}). 
In the next section, we give details of the expected nature of the 
mass-transfer modulation $\dot{M}_{tr}(\theta)$ at the orbital period.
    
\section{Orbital modulation of mass transfer}
\label{sec:masstransfer}

The problem of orbital modulation of mass transfer in eccentric orbits 
has been studied by a number of authors over almost three decades now, 
adopting various approaches appropriate for various aspects of
the problem they have studied. These aspects have covered a 
considerable range, from a scrutiny of the concept of the Roche lobe 
in an eccentric orbit (\cite{avni}), to a study of test-particle motion
through numerical integration of the restricted three-body problem
at or near periastron passage (\cite{lubow}), to explicit calculations of 
orbital phase-dependent flow through $L_1$ from a suitably-modeled
stellar envelope (\cite{joss} and references therein). 
For our purposes here, we have adopted the
results of the calculations described by Brown and Boyle (\cite{BB}, 
hereafter BB): these authors described the flow through $L_1$ from
the atmosphere of the lobe-filling companion with a scale height $H$ 
as a sort of nozzle flow through the inner Lagrangian point, 
integrating over a Maxwellian distribution of velocities
(characterized by thermal velocity scale $v_T$) for the stellar 
matter. Their final result for the rate of mass transfer as a function 
of the \emph{true anomaly} $\theta$ is given by:
\begin{equation}
\dot{M}_{tr}(\theta)=\dot{M_{0}}\frac{\gamma}{\gamma_{p}}\frac{1+e}
{1+e\cos\theta}\exp\left[ -\gamma\beta e\left(\frac{1-\cos\theta}
{1+e\cos\theta}\right)\right]. 
\label{eq:bbmdot}
\end{equation} 
In Eq.~(\ref{eq:bbmdot}), $e$ is the orbital eccentricity, and the 
dimensionless function $\gamma(\theta)$ is the ratio of the 
phase-dependent equivalent Roche-lobe radius $R(\theta)$ of the 
companion to the phase-dependent orbital distance $d(\theta)$ in
the eccentric orbit, $\gamma_{p}$ being the value of $\gamma$ at
periastron ($\theta=0$). From standard geometry of ellipses, 
$d(\theta)$ is given by:
\begin{equation}
d(\theta) = p{1+e\over 1+e\cos\theta},
\label{eq:ellipse}
\end{equation}
where $p\equiv a(1-e)$ is periastron distance. Finally, 
$\beta\equiv p/H$ is the the dimensionless scale-height parameter 
introduced by BB.   

It is convenient to work in 
terms of the ratio $\gamma$ as it varies relatively slowly with 
orbital phase (and is, in fact, independent of this phase for a
non-rotating companion: see below). The other properties of the 
binary system that $\gamma$ depends on are (a) the mass ratio 
$Q\equiv M_c/M_x$, $M_c$ being the mass of the low-mass 
companion, and (b) the rate of rotation $\Omega_c$ of the 
companion, usually expressed in units of the orbital angular 
velocity $\Omega_p$ at periastron as $\lambda\equiv\Omega_c/
\Omega_p$. The scale $\dot{M_{0}} = \sqrt{2\pi}\gamma_p pHv_T\rho_0$ 
of the mass-transfer rate in Eq.~(\ref{eq:bbmdot}) is set by the 
above velocity scale $v_T$, the scale-size $pH$ for the effective 
cross-section of the above ``nozzle'', and the basic density scale 
$\rho_0$ in the stellar atmosphere.  

Prescriptions for $\gamma$ have been given in the 1970s and '80s; 
we use here the generalized Joss-Rappaport (\cite{joss}) expressions 
adopted by BB, namely,
\begin{equation}
\gamma = A - B\log Q + C(\log Q)^2,
\label{eq:alpha1}
\end{equation}
where the coefficients in $\gamma$ are given by:
\begin{equation}
\left .
\begin{array}{l}
A=0.398-0.026K+0.004K^{3/2}\\
B=-0.264+0.052K-0.015K^{3/2}\\
C=-0.023-0.005K\\
\end{array} 
\right\rbrace 
\label{eq:alpha2}
\end{equation}
and the variable $K$ depends on the above rotation parameter 
$\lambda$ and the orbital phase as:
\begin{equation}
K=\lambda^2{(1+e)^4\over(1+e\cos\theta)^3}.
\label{eq:alpha3}
\end{equation}

From Eqs.~(\ref{eq:alpha1})-(\ref{eq:alpha3}), it is clear that,
for a non-rotating companion with $K=0$, $\gamma$ is independent
of the orbital phase, and depends only on the mass ratio $Q$. Thus,
for a given $Q$, $R(\theta)$ simply scales with $d(\theta)$ as the 
eccentric orbit is traversed. It is stellar rotation which modifies 
the Roche potential in such a way that this simple scaling is 
broken, and $\gamma$ depends on orbital phase. The phase-dependent 
factor in $K$ goes back to the original work of Avni (1976). In our
present work, we study the limits of (a) no stellar rotation, 
$\lambda=0$, and (b) synchronous stellar rotation, $\lambda=1$, to 
cover a range of possibilities (see below). The estimated accuracy 
in the above prescription for determining equivalent Roche-lobe radii
is $\sim 2$\%.

Detailed models with the mass transfer profile $\dot{M}_{tr}(\theta)$
given by Eq.~(\ref{eq:bbmdot}) are described below. 
From general considerations, it is clear that this profile 
peaks at the periastron and that the sharpness of the peak depends 
on the quantity $\gamma\beta e$. Since $\gamma\sim 1$, and typical 
values of $\beta$ for the current problem are in the range 
$10^{2}-10^{3}$ (BB), we see that the profile is expected to be
sharply peaked at the periastron even for realtively low values of 
eccentricity, such as $e\sim 0.2$.

\section{Viscous flow in accretion disks}
\label{sec:viscflow}
 
Matter transferred through $L_1$ into the Roche lobe of the neutron 
star first forms a ring around the neutron star, the radius $r_{ring}$
of this ring being related to the specific angular momentum 
$\ell_{tr}$ of the transferred matter as (\cite{pringle}):
\begin{equation}
r_{ring} = \ell_{tr}^2/(GM_x).
\label{eq:rout}
\end{equation}

Through effective viscous stresses, this ring spreads into an
accretion disk, wherein matter slowly spirals inward towards the
neutron star as the viscous stresses remove angular momentum from
it. The accretion disk extends from its outermost radius $r_{out}$ 
inward upto the magnetospheric boundary $r_m$, where the propeller 
torques expel the matter by depositing energy and angular momentum 
in it, as explained above.
 
The rate $\dot{M}$ at which the matter drifting radially inward
through the accretion disk arrives at $r_m$ depends, therefore,
both on the profile of mass supply $\dot{M}_{tr}$ at $L_1$, as 
described above, \emph{and} on the rate of viscous radial 
drift through the accretion disk, which occurs on a timescale 
$t_{visc}$.

In a quasi-steady state, the relation between the two profiles 
$\dot{M}(t)$ and $\dot{M}_{tr}(t)$ is of the form
\begin{equation}
\dot{M}(t)=\int_{t-N*P_{orb}}^t{\dot{M}_{tr}(t_0)f(\tau)dt_{0}},
\qquad {\rm where} \qquad \tau\equiv{t-t_0\over t_{visc}}.
\label{eq:mdotfin} 
\end{equation}
The convolution integral in Eq.~(\ref{eq:mdotfin}) describes the
viscous drift with the timescale $t_{visc}$ of the mass supplied 
to the disk at earlier times $t_0$ at the rate $\dot{M}_{tr}(t_{0})$,
as indicated above. In principle, the integral extends over all
previous times, but in practice it is sufficient to keep track of 
only about $N$ orbital periods in the past (as the lower limit of
integration indicates). This is so because of the rapid fall of
of the viscous-evolution profile $f(\tau)$ of the accretion disk 
at large values of $\tau$ (see below).

Viscous-evolution profiles have been calculated analytically and 
numerically at various levels of approximation by several authors 
(\cite{lynden}, \cite{lightman}). For our purposes here, we have 
adopted an analytic approximation of the generic form 
\begin{equation}
f(\tau)=\tau^{-n}\exp\left(\frac{-n}{\tau}\right)
\label{eq:profile}
\end{equation}
introduced and utilized by Pravdo and Ghosh (\cite{ghosh}, hereafter
PG). This reference has discussions of earlier analytical and
numerical investigations. The generic PG profile in 
Eq.~(\ref{eq:profile}) reaches its maximum at $\tau = 1$, and decays 
subsequently as $\tau^{-n}$. Clearly, therefore, most of the 
contribution to the above convolution integral comes from those 
orbital cycles which are closest to the earlier time $t_0 = t - 
t_{visc}$, and $N$ is determined by the sharpness of the fall of the 
profile, \ie, $n$. In our computations, we estimated the 
optimal values of $N$ by running test cases with increasing values 
of $N$ until the desired accuracy was obtained. For example, in the 
best-fit case reported below, we found that $N=9$ gave an accuracy
of $\approx 10$\%, while $N=15$ gave an accuracy of $\approx 1$\%.
Given the error bars on the data points in the observed light curve,
further accuracy was unnecessary.


\begin{figure*}
\centering
  \includegraphics [scale=0.3]{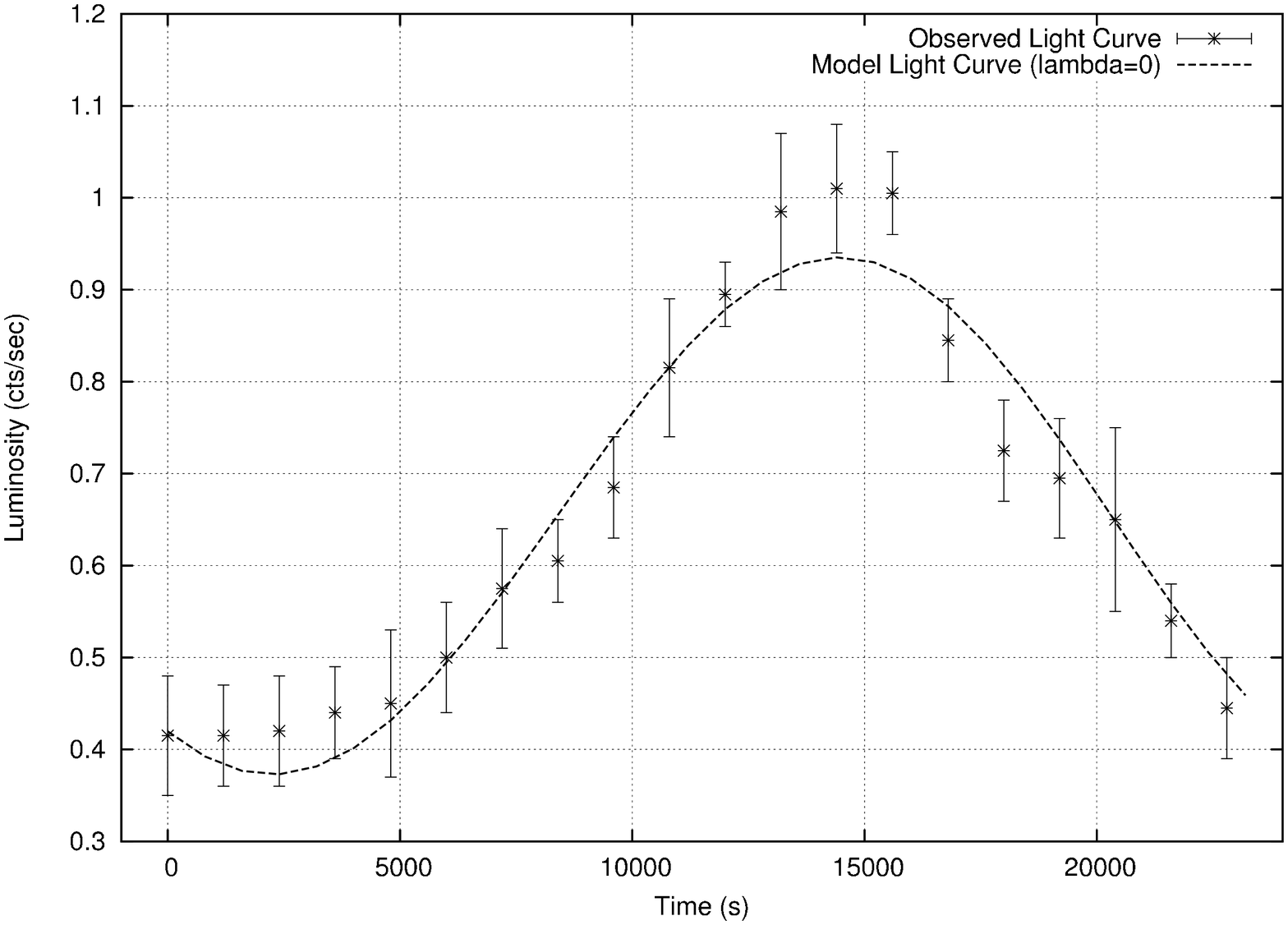}
   \includegraphics [scale=0.3]{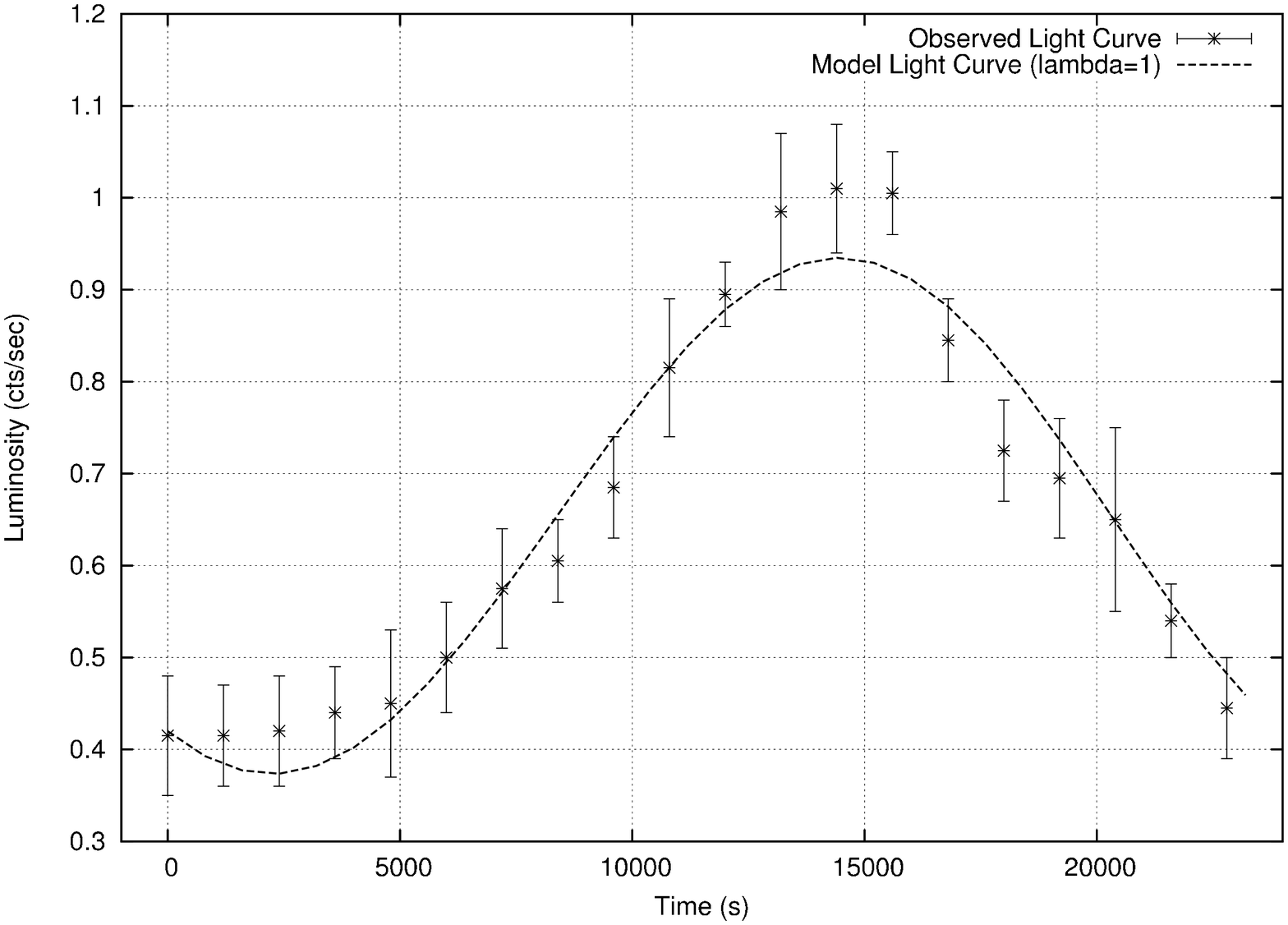}
   \caption{X-ray light curve of 1E. Shown is the observed light
curve from dL06, superposed on the (common) best-fit model light 
curve for IS-type and RUKL-type propellers.  
Left panel: model curve for $\lambda=0$ (nonrotating companion). 
Right panel: same for $\lambda=1$ (synchronously rotating companion).}
\label{modelfit}
\end{figure*} 

The following generic feature of viscous evolution of 
accretion disks is a key aspect of the phenomenon we are 
exploring here. Whereas the orbital modulation of the mass-supply 
rate $\dot{M}_{tr}(t)$ to the disk at its outer radius $r_{out}$ is 
expected to sharply peaked at periastron for typical values
of the scale height in the companion's atmosphere, as above, 
the viscous drift of matter through the accretion disk would 
decrease the sharpness of this modulation, since 
variations on timescales much shorter than $t_{visc}$ tend to be 
``washed out'' by viscous diffusion. This is what makes the 
orbital modulation of the mass-arrival rate $\dot{M}(t)$ at the 
disk's inner radius $r_m$ gentler, and therefore also the
modulation of the propeller luminosity $L(t)$, leading naturally
to light curves of the form observed in 1E. Quantitative details 
follow.

\section{Model light curves}

We constructed model light curves for 1E by combining the model
of mass transfer described in Sec.~\ref{sec:masstransfer} with   
that of viscous flow through the accretion disk described in 
Sec.~\ref{sec:viscflow}. We then fitted these models to the 
observed light curve of 1E in 2005 (dL06). 
The fitting parameters were ($\beta$, $e$), which come from the 
above BB mass-transfer model in elliptic orbit, and also 
($t_{visc}$, $n$), which come from the above PG parametrized 
description of viscous evolution of accretion disks. 
In this introductory work, we kept $\beta$
constant at a canonical value of $\beta = 10^3$ (BB), and varied
the parameters $e$, $t_{visc}$, and $n$ to obtain acceptable fits.
For the viscous timescale, we found it more convenient to work in 
terms of the ratio $\kappa\equiv t_{visc}/P_{orb}$ of this timescale
to the known period $P_{orb}\approx 6.67$ hr of the system, which
we of course identify with the orbital period in this model. The
ratio $\kappa$ is of immediate physical significance, since it
measures the relative importance of viscous diffusion to orbital
modulation in the system. For $\kappa\ll 1$, viscous 
diffusion would be so rapid as to enable the disk flow to adjust
to the orbital modulation of mass-supply rate, and flow-rate would 
essentially follow the supply rate. For $\kappa\gg 1$, on the other
hand, the viscous diffusion would be so slow as to wash out any
rapid variations in the mass-supply rate, and the  
modulation would be essentially determined by the disk viscosity.
As we see below, $\kappa$ values of a few seem to describe the 
1E system, indicating comparable importance of the two effects in
this system.  
  
\begin{table}[ht]
\caption{Best Fit Model Parameters: IS-type torque}  
\label{table:modelparamis}  
\centering                          
\begin{tabular}{| c | c | c |}        
\hline             
Parameter & Best fit value ($\lambda$=0) &  Best fit value
($\lambda$=1)\\    
\hline                        
   $\kappa$ & 2.6 & 2.6\\      
   viscous-profile index $n$ & 5.04 & 5.04 \\
   eccentricity  & 0.405 & 0.406 \\
\hline
   $\chi^2$ & 1.013 & 1.012    \\
\hline                                   
\end{tabular}
\end{table}

\begin{table}[ht]
\caption{Best Fit Model Parameters: RUKL-type torque} 
\label{table:modelparamrukl} 
\centering                          
\begin{tabular}{| c | c | c |}        
\hline             
Parameter & Best fit value ($\lambda$=0) &  Best fit value
($\lambda$=1)\\
\hline                        
   $\kappa$ & 2.6 & 2.6\\      
   viscous-profile index $n$ & 5.02 & 5.02 \\
   eccentricity  & 0.400 & 0.401 \\
\hline
   $\chi^2$ & 1.003 & 1.006    \\
\hline                                   
\end{tabular}
\end{table}

 We fitted model light curves corresponding to both IS-type and 
RUKL-type torques to the data on 1E, the best-fit values of the 
parameters being given in Table \ref{table:modelparamis} and Table
\ref{table:modelparamrukl}. In each case, we have considered both
a non-rotating secondary ($\lambda=0$) and a synchronously-rotating 
secondary ($\lambda=1$), as indicated. Note that the best-fit values
for the two types of torques are very close to each other, as may
have been expected. This is so because the closeness of the scaling
of $L$ with $\dot{M}$ between the two types, as discussed in 
Sec.~\ref{sec:propeller}, since only this aspect of the torque 
is relevant for fitting the \emph{profile} of the light curve. 
Other aspects, \eg, the fact that the RUKL-type propeller luminosity
is about three orders magnitude below the IS-type propeller 
lumnosity for identical vaues of $\mu$, $\dot{M}$ and $\omega$, have
important consequences elsewhere, as detailed in Sec.~\ref{sec:prop},
but not in this matter. Further, the absolute values of the observed
luminosities in the light curves are easily accounted for, \eg, by
having the stellar spin rate higher for RUKL-type torques by about
a factor of 10 than that for IS-type torques, and $\mu$, $\dot{M}$
identical for the two types, as the scalings in 
Eqs.~(\ref{eq:luminrukl}) and (\ref{eq:mdotprop}) show. This implies 
neutron-star spin periods in the range $P_{spin}\sim 0.01-0.1$ s, \ie,
the canonical range for propellers, for both type of torques, 
as explained in Sec.~\ref{sec:intro}.   

This closeness of best-fit parameters is reflected in the best-fit 
light curves, which are visually essentially identical for the two
types of torques. In Fig. \ref{modelfit}, we display this common
best-fit light curve, superposed on the data on 1E.

Our inferred best-fit value of $\kappa$ in the above tables indicates 
that the dominant contribution to the convolution integral described 
in the last section comes from the second and third orbits preceding
the time of observation. The corresponding viscous timescale 
$t_{visc}\approx 17.3$ hr is consistent with a rather thick disk with 
$h/r\sim 0.1-0.5 $ and a canonical value $\sim 0.1-1$ for the disk 
viscosity parameter (\cite{shakura}). This seems consistent with 
the results of the RUKL numerical simulations. Note also that
the best-fit value of the viscous-profile index $n$
is consistent with the range of values $n\sim 4-5$ generally 
expected for neutron-star systems, as per the discussion given 
in PG. Indeed, we found that values of $n$ in the above range
generally worked for the 1E system. Regarding the orbital 
eccentricity $e$, the best-fit values are as given in the tables,
and we found that values of the eccentricity $e$ in the range 
$\sim 0.35-0.45$ genearlly worked for the 1E system: we discuss 
this in the next section. It is clear, therefore, that the model 
explored in this paper can account quantitatively for the observed 
1E light curve in 2005, for both IS-type and RUKL-type torques. 
We discuss in Sec.~\ref{sec:ldepltcurve} possible reasons for the 
apparently different, ``jagged'' light curve hinted at by the 2001 
observations of this system (dL06).

\section{Formation \& evolution of prototype systems}
\label{sec:formation}   

As indicated in Sec.~\ref{sec:intro}, we are exploring in this work
a model for systems like 1E wherein the binary system of a He-star 
and a low-mass
star (left after completion of the CE evolution phase in which the
extensive envelope of the evolved primary has been expelled and its
He-core left behind) produces the pre-LMXB when the He-star explodes
in a supernova (SN), leading to a newborn neutron star with a low-mass
companion. Essential features of the formation and subsequent 
evolution of such systems are, therefore, essential components of
this model. We now discuss these features in brief, considering in
this section first the immediate post-SN status of the system, and
then the evolution of this system with the low-mass companion in an 
eccentric orbit at or near the point of Roche-lobe contact at
periastron, producing a system like 1E where orbitally-modulated mass 
transfer proceeds through the inner Lagrangian point, and the newborn, 
fast-spinning neutron star is operating in the propeller
regime, expelling this matter instead of accreting. Subsequently, 
we summarize further evolution of such systems.

\subsection{Immediate post-SN systems}
\label{sec:postsn}  

A major question that concerns us here is the expected eccentricity
of systems formed by the SN in the above scenario, since this eccentricity
is crucial for the proposed mechanism. Qualitatively, it is obvious 
that the immediate post-SN system is almost guaranteed to be highly 
eccentric, as the mass loss from a typical pre-SN system of, say, a  
$3\Msun$ He-star and a $M_c=0.4\Msun$ low-mass companion (see
below) in forming the post-SN system of $M_x=1.4\Msun$ neutron star
with its $M_c=0.4\Msun$ low-mass companion is $1.6\Msun$, which
is close enough to maximum allowed value of mass loss ( = half of
the initial total mass of $3.4\Msun$ for zero kick velocity) to 
ensure that the post-SN orbit would be very eccentric. We shall use 
these values for the stellar masses throughout the rest of this paper.           

To see this quantitatively, we can adapt the extensive calculations of
Kalogera, who computed the probability of the formation of X-ray
binaries as a funtion of orbital parameters (\cite{vicky}). In the 
following, we shall use the same masses for the pre- and post-SN
system as given above for illustrative purposes. The probability
density from Kalogera's work is: 
\begin{eqnarray}
G(\alpha,e) & = &\left(\frac{\zeta}{2\pi\xi^2}\right)^{3/2}
\frac{2\pi e}{\left[\alpha (1-e^2)\right]^{1/2}}\times\\\nonumber
&&\left[\left(\alpha-\frac{1}{1+e}\right)
\left(\frac{1}{1-e}-\alpha\right)\right]^{-1/2}\times\\\nonumber
&&\exp\left[-\frac{1}{2\xi^2}\left(\zeta~\frac{2\alpha-1}
{\alpha}+1\right)\right]~I_o(z)
\label{eq:probdens}
\end{eqnarray}
Here,
\begin{displaymath}
z~\equiv~
\frac{\left(\zeta~\alpha~(1-e^2)\right)^{1/2}}{\xi^2},
\end{displaymath}
and $I_o$ is the modified Bessel function of zeroth order. Further, 
$\alpha$ is the ratio of semimajor axes of the pre- and post-SN
orbits, $\zeta$ is the ratio of the total mass of the post-SN
binary to that of the pre-SN one, and $\xi \equiv \sigma /V_r$, 
$\sigma$ being the velocity dispersion in the SN kick-velocity, 
and $V_r$ the orbital velocity of the exploding star relative 
to its low-mass companion just before the SN (\cite{vicky}).      

\begin{figure}[h]
 \centering
  \includegraphics[width=9cm]{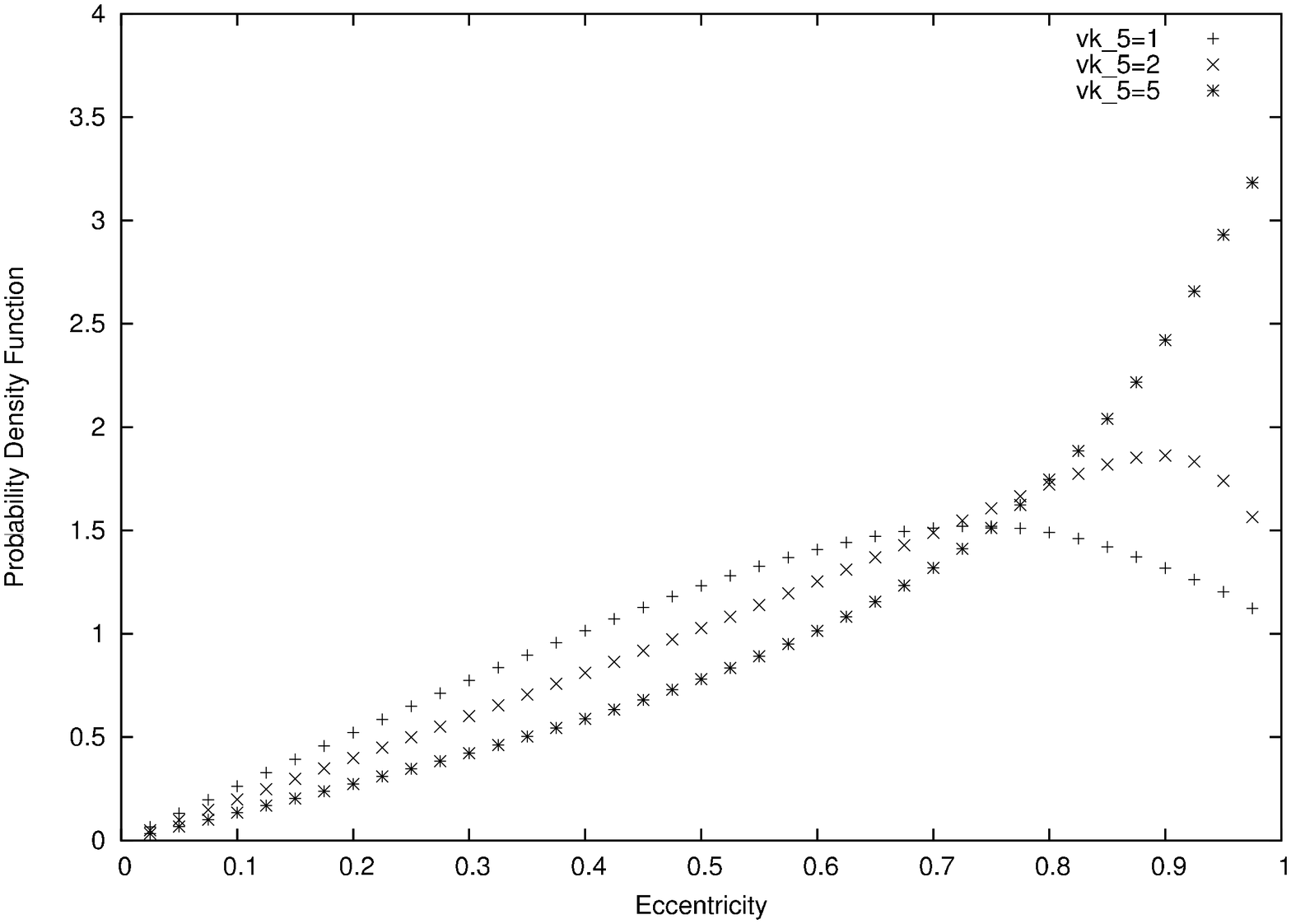}
  \caption{Formation probability-density ${\cal G}(e)$ of 
immediate post-SN binaries as a function of eccentricity $e$
for various values of the dispersion $\sigma$ in the SN kick 
velocity (see text). Curves labeled by the value of $vk_5 = \sigma$ 
in units of 100 km s$^{-1}$. Each curve so normalized that 
$\int{\cal G}(e)de=1$.}
  \label{normfe}
\end{figure}

In the problem we are studying here, the semimajor axis of the 
post-SN binary is determined by Kepler's third law from our 
assumed stellar masses above, and the known orbital period of 1E. 
However, when there is a kick associated with the SN, the inferred 
semimajor axis of the pre-SN binary is not determined uniquely by 
the semimajor axis and the eccentricity of the post-SN binary: 
rather, there is a range of values corresponding to the range of 
the kick-velocity. Thus, there is a range in the values of $\alpha$: 
it is well-known that the allowed range for $\alpha$ is limited 
from $1/(1+e)$ to $1/(1-e)$, these limits being first identified by 
Flannery and van den Heuvel (1975). Thus, for our purposes, it is 
aprropriate to integrate $G(\alpha,e)$ over the above allowed range 
of $\alpha$, and display the resultant probability density 
${\cal G} (e)\equiv\int G(\alpha,e)d\alpha$ as a function of the 
eccentricity $e$. We show this in Fig.~\ref{normfe} for various 
typical values of $\sigma$ as indicated. In this figure, we have 
used the symbol $vk_5$ there to denote $\sigma$ in units of $10^5$ 
m s$^{-1}$ = 100 km s$^{-1}$, the typical scale for the SN kick 
dispersion, and we have normalized the 
probability density ${\cal G}(e)$ so that 
$\int{\cal G}(e)de=1$ in each case. As explained above, the closeness 
of the value of $\zeta\approx 0.53$ in this typical case to its lower 
limit for no binary destruction in the SN (this limit is 0.5 for zero 
kick velocity) ensures that the probability peaks at a high value of 
$e$, as Fig.~\ref{normfe} shows. It is clear, therefore, that such a 
pre-LMXB would generically have a considerable eccentricity at the 
time of its formation in the SN.    

\subsection{Tidal-evolution phase of pre-LMXBs}
\label{sec:tidal}

The above newly-formed pre-LMXB undergoes tidal evolution,
wherein three simultaneous processes occur, namely, (1) tidal 
circularization, \ie, decrease in the orbital eccentricity $e$, 
(2) tidal orbit-shrinkage or \emph{hardening}, \ie, decrease in the 
orbital semimajor axis $a$, and (3) tidal synchronization, whereby 
the rotation frequency $\Omega_c$ of the low-mass companion approaches 
the orbital angular frequency $\Omega\equiv 2\pi/P_{orb}$. These processes 
happen through tidal torques, and their quantitative descriptions 
pioneered by Zahn (1977, 1978) are widely used for calculations: we 
use them here, as have P08. Complete equations 
are given in Zahn (1977), and an Erratum was published by Zahn
(1978). We have found a further algebraic or transcription error in
the original paper, which we describe below, and which seems to have
gone unnoticed so far. 

Complete formulations for the rates of change of $e$, $a$, and 
$\Omega_c$ are given in Zahn (1977), but for our work here we shall
utilize a widely-used simplification which comes naturally out of 
these formulations, namely, that the timescale for tidal synchronization
comes out to be much shorter than that for tidal circularization and
tidal hardening (see, \eg, \cite{MM}, P08). This is appropriate,
since we shall be interested in this work only in phenomena which
occur on the timescales of tidal circularizatuion or longer. Under
such circumstances, we can look upon the system as being roughly 
synchronous at all times, and describe tidal circularization and 
tidal hardening respectively by Zahn's (1977) equation (4.7) and the
appropriately simplified (\ie, synchronized) version of Zahn's
equation (4.3), thereby obtaining:

\begin{equation}
-{1\over e}{de\over dt} = {63\over 4}{k_2\over t_F}q(1+q)
\left({R\over a}\right)^8 ,
\label{eq:tidcirc} 
\end{equation}   

and

\begin{equation}
-{1\over a}{da\over dt} = 114{k_2\over t_F}q(1+q)
\left({R\over a}\right)^8e^2 .
\label{eq:tidhard} 
\end{equation} 
In equations (\ref{eq:tidcirc}) and (\ref{eq:tidhard}), $q\equiv 1/Q$
in terms of the mass ratio $Q\equiv M_c/M_x$ defined above in 
Sec~\ref{sec:masstransfer}, $k_2$ is the apsidal motion constant for
the low-mass companion, and $t_F$ is the ``friction time'' of
Zahn (1977), which, for stars with convective envelopes (as in the
present case) is given by Zahn's (1977) pioneering prescription of 
the turbulent eddy-viscosity timescale $t_{EV}$:

\begin{equation}
t_F \sim t_{EV} = (M_cR_c^2/L_c)^{1/3} .
\label{eq:eddyvisc}
\end{equation} 

Equations (\ref{eq:tidcirc}) and (\ref{eq:tidhard}) describe 
simultaneous tidal circularization and hardening of close binaries,
but before presenting our results, we need to correct two errors
related to them. First, if we define a circularization timescale 
$t_{circ}\equiv -e/(de/dt)$ in the usual way, we get from 
eq.~(\ref{eq:tidcirc}) the result:

\begin{equation}
t_{circ} = {4\over 63}{1\over k_2q(1+q)}\left(M_cR_c^2\over 
L_c\right)^{1/3}\left({a\over R}\right)^8 ,
\label{eq:tcirc} 
\end{equation}
 which would be identical to Zahn's (1977) equation (4.13), except
that the factor of 4 on the right-hand side is missing in Zahn
(1977). Unfortunately, this error has propagated over the years
into numerous papers, \eg, in P08, in their 
equation (2)\footnote{Because of this, the parameters adopted for
1E by P08 and by ourselves in this work actually
give $t_{circ}\sim 10^4$ yr. In our work, we have used the values
of the apsidal-motion constant $k_2$ given by Landin \etal~(2009).} 
We have corrected this now. Secondly, in an erratum  
published in 1978, Zahn corrected a few other (generally smaller)
numerical errors, of which the one relevant to our work is that
the numerical coefficient on the right-hand side of our 
eq.~(\ref{eq:tidcirc}) should be 21 instead of 63/4. In all 
calculations reported here, we have made these corrections.  

\begin{figure}[h]
 \centering
  \includegraphics[width=9cm]{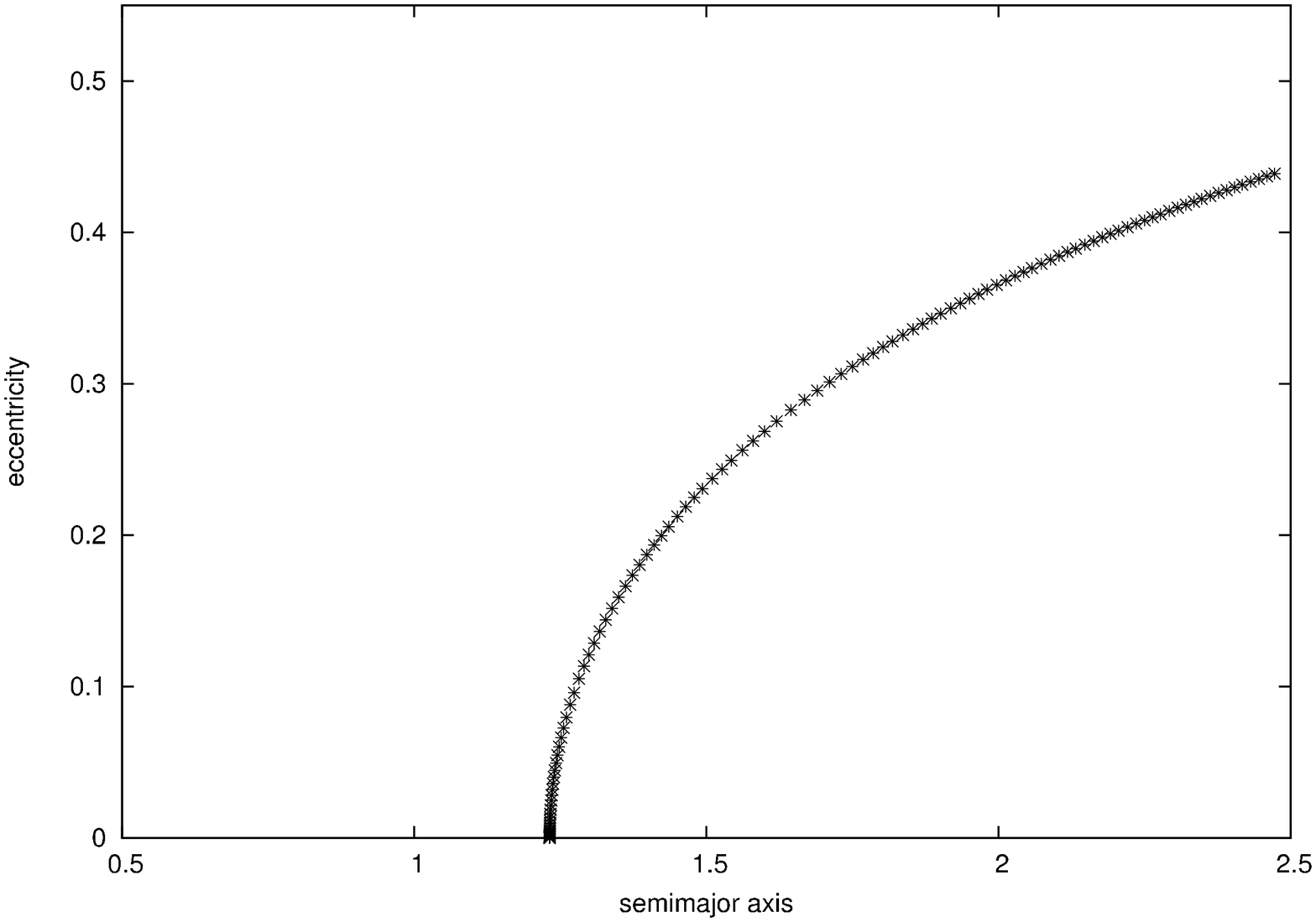}
  \caption{Tidal evolution of a prototype 1E-like system in the
$e$ vs. $a$ plane. Semimajor axis $a$ in units of solar radius. 
Note the ``cut off'' like approach to the circularization 
point (see text).}
  \label{cutoff}
\end{figure}

We have integrated eqs.~(\ref{eq:tidcirc}) and (\ref{eq:tidhard})
numerically for close binary systems like 1E, with values of 
initial post-SN semimajor axes and eccentricities, $a_i$ and $e_i$, 
chosen over a range of plausible values for such systems. We find 
that, in all cases, the systems circularize and harden in a way 
that, in the ($e$ vs. $a$) plane, the circularization point is 
approached in a ``cut off'' like manner. This is shown in  
Fig.~\ref{cutoff} for a possible prototype 1E-like system, so
chosen that the parameters of it evolve to those roughly
corresponding to 1E in $\sim 2000$ years. This cut-off approach  
is similar to what Meibom and Mathieu (2005) found. Of course, 
our detailed shape is slightly different from that of these authors, 
since they fitted their results to an assumed parameterized 
distribution shape applicable to observations on a collection of 
``normal'' binaries.
These details wil be given in a separate publication. For our
purposes here, we note that the total time $\tau_{circ}$ taken to
reach this circularization point (\cite{MM}) can be expressed
roughly as:
\begin{equation}
\tau_{circ} \approx \tau_0\left(a_i\over R_c\right)^8e_i^{-2.55},
\label{eq:taucirc}     
\end{equation}
where $a_i$ and $e_i$ are the initial semimajor axis and 
eccentricity of the immediate post-SN orbit, and the scale
parameter $\tau_0$ is given by:

\begin{equation}
\tau_0 \approx {1\over 2k_2q(1+q)}\left(M_cR_c^2\over 
L_c\right)^{1/3}.       
\label{eq:tauzero}
\end{equation}
Equation (\ref{eq:taucirc}) is a rough analytic fit to the 
mumerical results, adequate for our purposes. Note that the 
scale parameter $\tau_0$ depends on the companion mass $M_c$, its
value being $\tau_0\approx 1$ yr for the inferred companion
mass of 1E.

It is clear from eq.~(\ref{eq:taucirc}) that circularization is 
faster for orbits which are born more compact and more 
eccentric. The scaling with $a$ is straightforward from the 
above equations of tidal evolution; the scaling with $e$ is
more complicated (although inspection of the same equations
gives some clue), involving details of the numerical solution.
  
The lifetime $\tau_{circ}$ of the eccentric phase of the pre-LMXB
is obviously also the lifetime of its orbital-modulation phase
which we are investigating in this work. The sensitive dependence
of this lifetime on the initial post-SN orbital parameters and
the companion mass (through the scale parameter $\tau_0$ and due
to the mass-dependence of $R_c$ in eq.[\ref{eq:taucirc}]) makes
for a wide range of possible values of this lifetime, $\sim 10^3
- 10^8$ years. 

\begin{figure}[h]
 \centering
  \includegraphics[width=9cm]{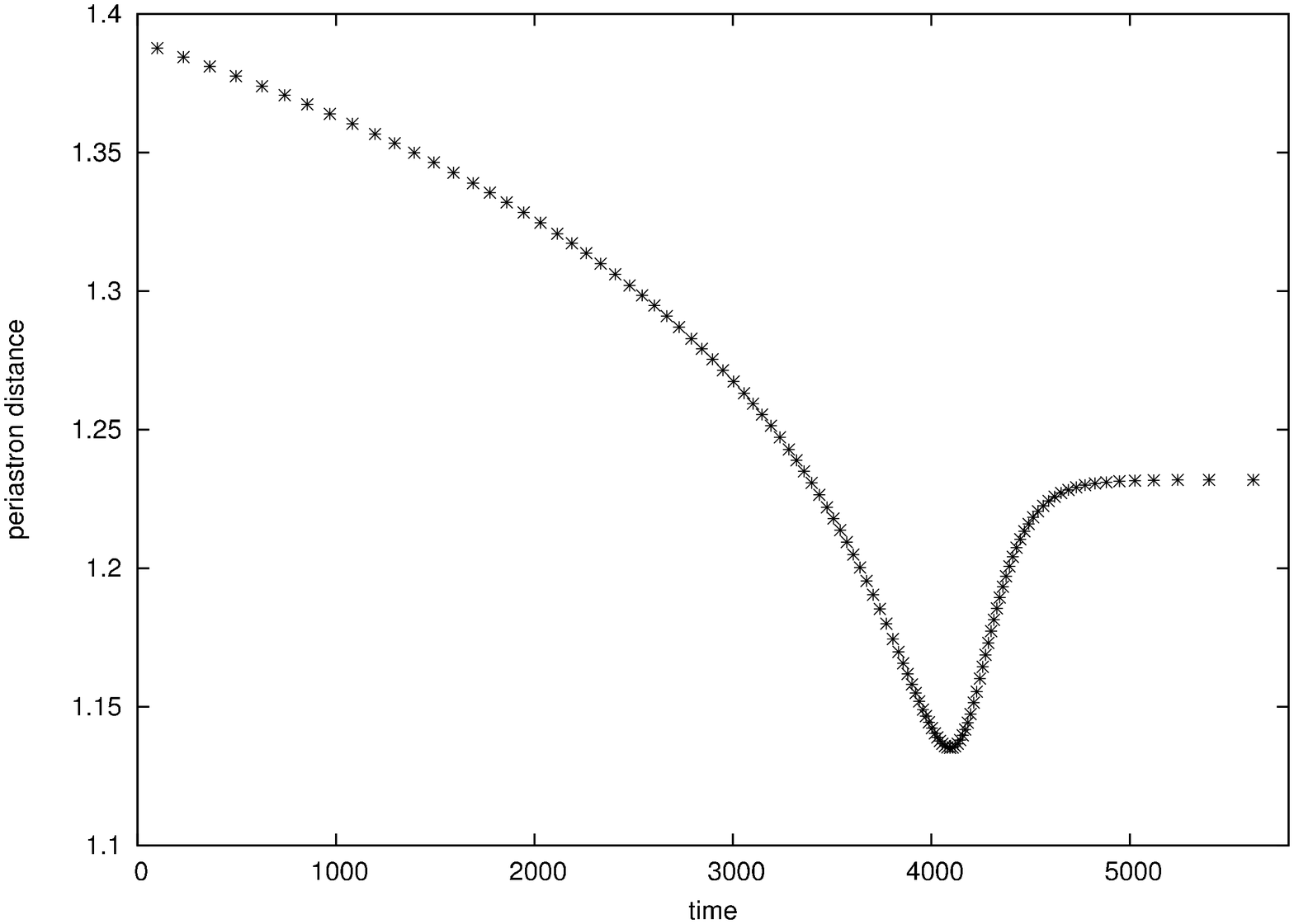}
  \caption{Evolution of periastron distance $p=a(1-e)$ during 
tidal evolution of a prototype 1E-like system (see text). Shown 
is $p$ in units of the solar radius vs. time in years.}
  \label{rl}
\end{figure}

A crucial point is, of course, that if the companion is at or 
close to filling its Roche lobe at periastron in the post-SN 
orbit, it must remain so throughout most of this eccentric phase
in order for the scenario to be self-consistent. The size of the
Roche lobe at periastron is simply $p=a(1-e)$ multiplied by a
well-known function of the mass ratio $q$. Since the latter 
does not change significantly during this phase, we need only
study the evolution of the former. Our integration of the 
tidal-evolution equations show that, while $a$ and $e$ both 
decrease during this phase, $p=a(1-e)$ decreases slowly through 
most of this phase, reaching a minimum and increasing thereafter
at late stages. This is shown in Fig.~\ref{rl} for the prototype
1E-like system displayed in Fig.~\ref{cutoff} (see above). 
Thus, if the companion is initially at or close 
to filling its Roche lobe at periastron, it will remain so over 
most of this phase, and if it is inside its Roche lobe initially,
it is likely to fill its Roche lobe later during this phase. It 
is also seen that Roche-lobe contact ends at the last parts
of this phase (when the orbit is nearly circular), since $p$ 
increases and becomes roughly constant there.

Thus, this tidal-evolution phase is a rough measure of the 
lifetime of Roche-lobe contact and orbital modulation of the 
propeller output. After this, the pre-LMXB becomes detached, and
remains so until angular-momentum loss through gravitational
radiation and/or magnetic braking brings it back to Roche-lobe
contact on a long timescale of $10^8-10^9$ yrs. We discuss this
phase below in Sec.~\ref{sec:recontact}. 

\subsection{Duration of propeller phase}
\label{sec:prop}

When the above tidal-evolution phase ends, is the neutron star
still operating in the propeller phase? To answer this question, 
we consider the spindown of the neutron star from an initial 
spin period $P_{spin}^i$ to a final, longer spin period
$P_{spin}^f$ under the action of the propeller torque given by
either the IS-type torque (Eq.~[\ref{eq:sstorque}]) or the 
RUKL-type torque (Eq.~[\ref{eq:rukltorq}]). In each case, 
this spindown is decsribed by
\begin{equation}
{\dot{\omega}\over\omega} = {N\over I\omega} = {1\over t_{prop}}
\label{eq:spindown}
\end{equation}
where $I$ is the moment of inertia of the neutron star and 
$t_{prop}$ is the propeller spindown timescale. 

First consider IS-type torques, for which $t_{prop}$ is 
given by
\begin{equation}
t_{prop} = 3\sqrt{2}{(2GM_x)^{2/7}I\over\mu^{8/7}\dot{M}^{3/7}}
\approx 2.5\times 10^5 \dot{M}_{14}^{-\frac{3}{7}}
\mu_{30}^{-\frac{8}{7}}m_{x}^{\frac{2}{7}} I_{45}\:\:{\rm yr}, 
\label{eq:proptimeis}
\end{equation}     
where $I_{45}$ is $I$ in units of $10^{45}$ gm cm$^2$, and other
units are as before. Equation~(\ref{eq:spindown}) can be 
integrated readily in this case, the total spindown time $\tau_{prop}$ 
from $P_{spin}^i$ to $P_{spin}^f$ being:
\begin{equation}
\tau_{prop} = 2.303t_{prop}\log(P_{spin}^f/P_{spin}^i).
\label{eq:propduris1}
\end{equation}
As discussed earlier, the ratio $P_{spin}^f/P_{spin}^i$ is believed
to be in the range $10-100$ (\cite{ghoshprop} and references therein), 
and its exact value does not matter because of the logarithmic 
dependence. On taking $m_X=1.4$ and the corresponding moment of
inertia for a standard modern EOS, we arrive at 
\begin{equation}
\tau_{prop} \approx 2\times 10^6 \:\:{\rm yr} 
\label{eq:propduris2}
\end{equation}
for canonical values of $\dot{M}$, $\mu$ and $I$.

Now consider RUKL-type torques, for which $t_{prop}$ is given by
\begin{eqnarray}
t_{prop} & = & t_0(P_{spin}/0.01 {\rm s}),\:\:\: {\rm where}\nonumber\\  
t_0 & \approx & 2.3\times 10^7 \dot{M}_{14}^{-\frac{3}{8}}
\mu_{30}^{-1.1}I_{45}\:\:{\rm yr}. 
\label{eq:proptimerukl}
\end{eqnarray}
Equation~(\ref{eq:spindown}) can be integrated readily in this case
also, the total spindown time $\tau_{prop}$ from $P_{spin}^i$ to 
$P_{spin}^f$ being:
\begin{equation}
\tau_{prop} = t_0(P_{spin}^f - P_{spin}^i)/(0.01 {\rm s})
\approx t_0(P_{spin}^f/0.01 {\rm s}),
\label{eq:propdurrukl1}
\end{equation}
the second equality in the above equation coming from the fact that
the ratio $P_{spin}^f/P_{spin}^i$ is believed to be in the range
$10-100$, as indicated above. The numerical value of $\tau_{prop}$
in this case is thus
\begin{equation}
\tau_{prop} \approx 2.3\times 10^7\dot{M}_{14}^{-\frac{3}{8}}
\mu_{30}^{-1.1}I_{45}(P_{spin}^f/0.01 {\rm s})\:\:{\rm yr},  
\label{eq:propdurrukl2}
\end{equation}
which implies that, for canonical range $P_{spin}^f\sim 0.1-1$ s,
as indicated in Sec.~\ref{sec:propeller}, we arrive at
\begin{equation}
\tau_{prop} \approx 2\times 10^8 - 2\times 10^9\:\:{\rm yr} 
\label{eq:propdurrukl3}
\end{equation} 
for canonical values of the variables $\dot{M}$, $\mu$ and $I$. 

In comparing the total spindown times $\tau_{prop}$ given by the 
two types of torques, we notice that the time taken by the 
RUKL-type torque is 2-3 orders of magnitude longer than that 
taken by the IS-type torque for identical values of $\dot{M}$, 
$\mu$ and $I$. This reflects the relative weakness of the 
RUKL-type torque discussed in Sec.~\ref{sec:propeller}.
Next, comparing the values of $\tau_{prop}$ given by the above two 
types of propeller torques with the lifetime $\tau_{circ}$ of the
eccentric phase given in the previous section, we reach the  
following conclusions. For the IS-type torque, we find that, over 
most of the parameter space, the neutron star would still be in the 
propeller phase at the end of the above tidal-evolution phase of 
the binary. For the RUKL-type torque, we find that this conclusion 
is valid over the entire parameter space. Thus, the RUKL-type 
torque makes the conclusion stronger.  

As shown above, the companion has moved out of Roche-lobe contact 
by the time that the tidal-evolution phase of the binary reaches
conclusion, so that mass transfer stops, and so does the propeller 
action and its consequent soft X-ray production. Accordingly, 
throughout this first Roche-lobe contact phase, we expect the 
system to be in the propeller phase.    

\subsection{Re-contact with Roche lobe \& LMXB phase}
\label{sec:recontact}

After orbit circularization and the loss of its first Roche-lobe 
contact, as described above, the pre-LMXB thus ceases to be an 
X-ray source. But its orbit shrinks (\ie, the binary \emph{hardens}) 
on a long timescale ($\sim 10^8-10^9$ yr)
due to two mechanisms of angular momentum loss from the 
system, \viz, graviational radiation and magnetic braking 
(\cite{ghoshbook} and references therein). These are the standard
mechanisms through which short-period pre-LMXBs are believed to 
harden, until Roche-lobe contact is regained and mass transfer
restarts. But the transferred mass is now accreted by the neutron 
star, because its spin has been slowed down sufficiently over
this long time that it acts as an accretor and not a propeller
at the (large) mass-transfer rates that occur at this second 
Roche-lobe contact in the circularized binary. The system thus 
turns on as a canonical LMXB now, emitting strongly ($L\sim 10^{37}
-10^{38}$ erg  s$^{-1}$) in the canonical X-ray band characteristic
of emission from the neutron-star surface, rather than the soft
X-ray band characteristic of propeller emission from the vicinity
of the magnetospheric boundary.

The timescale $t_{GR}$ of orbit shrinkage due to gravitational 
radiation is given by (see, \eg, Faulkner 1971, Banerjee \& Ghosh
2006):
\begin{equation}      
t_{GR} \approx 2\times 10^9{m_T^{1/3}\over m_xm_c}
\left({P_{orb}\over 6^h.7}\right)^{8/3}\:\: {\rm yr}
\label{eq:tgr}
\end{equation}
where $m_T\equiv m_x+m_c$, and all masses are in solar units.
In this equation, we have scaled $P_{orb}$ to the value for 1E,
and substitution of the masses we have used above for this system
gives $t_{GR}\approx 4\times 10^9$ yr. Generally, 1E-like systems 
with shorter periods and/or somewhat different companion masses
will have $t_{GR}\sim 10^8-10^9$ yr. Magnetic braking is believed
to be comparable or weaker in strength to shrinkage by gravitational
radiation at these orbital periods, so that the above estimate is
a reasonable one for the 1E-type systems we have in mind here.

Thus, the system become a canonical, bright LMXB with a circular
orbit and $P_{orb}$ in the range of, say, 2 -- 10 hours. It is 
well-known that systems with $P_{orb}$ exceeding about 12 hours
cannot come into Roche lobe contact by the above orbit-shrinkage
mechanisms, since the time requied would exceed the Hubble time,
as eq.~(\ref{eq:tgr}) readily shows. However, these long-period 
systems also come into Roche-lobe contact eventually, as the
low-mass companion completes its main-sequence evolution and
expands. These systems thus also become canonical long-period
LMXBs with circular orbits. The lifetime of this standard, bright 
LMXB phase is $t_{LMXB}\sim 10^8-10^9$ yr.      

\section{Discussion}
\label{sec:discuss}

In this work, we have explored a pre-LMXB model of 1E, wherein the
eccentric orbit of the very young pre-LMXB causes an orbital 
modulation in the mass-transfer rate, and the newborn, fast-rotating
neutron star operates in the propeller regime, the propeller 
emission in soft X-rays following the above modulation after 
viscous smoothening in the accretion disk. In this section, we first
discuss first some essential spectral and luminosity-dependent 
features of 1E, and their connections with corresponding features 
in old, low-mass, soft X-ray transients (SXRTs) in their low/quiescent 
states, the prime example of this class being Aquila X-1 (\cite{campana}). 
Note that the well-known transient accretion-powered millisecond 
pulsar SAX J1808.4-3658 also shows a similar behavior (\cite{stella}). 
In these classes of low-mass X-ray binaries with old neutron stars,  
the neutron star is thought to operate in the propeller regime when 
the sources are in their low/quiescent states during decays of their 
outbursts. We then compare our model with the magnetar model 
which has been proposed recently for 1E (P08), 
and discuss how distinction between the two kinds of models might
be attempted in future. Finally, we summarize our conclusions.      

\subsection{X-ray spectra}
\label{sec:spectra}

The XMM-Newton/EPIC (0.5-8 keV) X-ray spectra of 1E have been 
described by dL06. The time-averaged spectra from
the 2005 low-state observations, when the source luminosity was
$L\sim 10^{33}$ erg s$^{-1}$, can be fitted by a two-component model 
consisting of a blackbody (BB) of temperature $kT_{bb}\sim 0.5$ keV and an 
equivalent blackbody radius $R_{bb}\sim 0.6$ km, plus a power-law (PL)
of index $\Gamma\sim 3$, with $\sim 70$\% of the total flux coming
from the blackbody component. 
Alternatively, the second component can also be a 
blackbody with a higher temperature. A re-analysis of the earlier 2001 
XMM-Newton data, when 1E had a higher luminosity (by a factor $\sim 6$),     
yielded a similar two-component (BB+PL) model with essentially the 
same blackbody temperature $kT_{bb}$ and power-law index $\Gamma$, but 
a larger equivalent blackbody radius $R_{bb}\sim 1.3$ km, and a higher
contribution from the PL component (the blackbody contribution was 
$\sim 50$\% of the total flux as opposed to the above $\sim 70$\%),
which made the overall spectrum harder (dL06).  

We stress the remarkable similarity of the above observations with 
those of the spectra of SXRTs in their low/quiescent states (when
the neutron stars in them are believed to be functioning in the 
propeller regime), taking the well-known source Aquila X-1 as the
example. A detailed analysis of the BeppoSAX observations of Aquila X-1   
in 1997 (\cite{campana}) has yielded the following results. At the
lowest state, with source luminosity $L\sim 0.6\times 10^{33}$ erg
s$^{-1}$, the (BB+PL) fit had a BB of temperature $kT_{bb}\sim 0.3$ keV and an 
equivalent blackbody radius $R_{bb}\sim 0.8$ km, plus a power-law (PL)
of index $\Gamma\sim 1$, with $\sim 60$\% of the total flux coming
from the blackbody component. As the luminosity increased by a 
factor $\sim 150$ to $L\sim 9\times 10^{34}$  erg s$^{-1}$, the    
(BB+PL) fit yielded a BB of temperature $kT_{bb}\sim 0.4$ keV and an 
equivalent blackbody radius $R_{bb}\sim 2.6$ km, plus a power-law (PL)
of index $\Gamma\sim 1.9$, with $\sim 20$\% of the total flux coming
from the blackbody component. Remembering that the total range of
luminosities in these Aquila X-1 low-state observations during outburst
decay is roughly $10^{33}-10^{35}$  erg s$^{-1}$ (\cite{campana}), 
essentially identical to that of the 1E observations reported by dL06, 
the correspondence is very suggestive.

SXRTs are believed to be old systems with a neutron star and a
low-mass companion in a close circular orbit, undergoing outbursts due to 
instabilties either in the accretion disk or in the mass supply from
the low-mass companion. In their low/quiescent states during decays 
of outbursts, the fast-spining neutron star (spun up by accretion as 
per standard LMXB scenario) is believed to operate in the propeller
regime. What we suggest in this work is that 1E-like systems are
very young systems in the same regime: the young systems can show 
orbital modulation because of the orbital eccentricity, while the old 
systems are in circular orbit and cannot show such orbital modulation.
However, the spectral signatures are very similar at similar 
luminosities, which supports our basic suggestion. We note that
the timescales associated with 1E outburst appear to be $\sim 2-3$ 
years while those associated with Aquila X-1 outbursts appear to be 
$\sim 30-70$ days. It is possible that the basic phenomenon is rather 
similar in the two cases, and that the difference in detail is caused 
by the fact that accretion onto the neutron-star surface (with 
attendant high luminosities and hard X-ray spectra) does occur at
the high states during the outbursts for old systems like Aquila X-1, 
but not for young systems like 1E.  

A comprehensive theory of the emission spectra of propeller sources 
appears to be lacking, though Illarionov and co-authors have studied 
some effects of Comptonization in propellers in wind-accreting massive 
X-ray binaries (\cite{ik,iik}). Attempts at constructing such a
theory for propellers in pre-LMXBs and in old LMXBs in low/quiescent 
states is clearly beyond the scope of this paper, and we shall confine 
ourselves here to the comment that the importance of Compton heating,
considered in the above works on propellers in massive binaries, is 
also likely to be crucial for the systems we are focusing on in this
work, as the observed power-law tails in the spectra at low 
luminosities suggest. These tails are particularly prominent in the 
low-state spectra of Aquila X-1 (\cite{campana}).      

\subsection{Luminosity dependence of light curve}
\label{sec:ldepltcurve}

dL06 have compared the 1E light curve in the 
2005 low-state observations with that during the 2001 observations
when the source luminosity was a factor $\sim 6$ higher. While 
the former light curve is relatively smooth with some cycle-to-cycle   
variations, the latter one shows more complex, somewhat ``jagged'' 
structure, with an occasional dip. Further, the pulsed fraction 
decreses from $\sim 43$\% to $\sim 12$\% as the luminosity 
increases. We discuss qualitatively how such features may arise.
First, a propeller system is inherently more fluctuating than an
accreting system, because of a variety of fluctuations possible
at the site of shock-heating and outflow. As mass-supply rate 
through the accretion disk increases, these fluctuations may 
increase, causing more complex profiles. Secondly, accretion disks
in low-mass systems like LMXBs and CVs are thought to develop 
structures at their outer edges, which obscure emission from the
compact object, and lead to dips. If these obscuring structures
increase in size as mass-supply rate through the accretion disk 
increases, this would provide a natural explanation for the
above appearance of the dips. Thirdly, as the mass-arrival rate
$\dot{M}$ at $r_m$ increases, $r_m$ decreases (see Sec.~\ref{sec:propeller}), 
matter at the magnetospheric boundary becomes hotter, and the propeller 
becomes less supersonic, ultimately becoming subsonic. Now, it is 
well-known that the subsonic propeller torque 
$N_{sub}\sim\mu^2\Omega_s^2/GM_x$ is \emph{independent} of $\dot{M}$
(see \cite{ghoshprop} and references therein), and so will not
follow the modulations of $\dot{M}$. Hence, as $\dot{M}$ and $L$ 
increase, the following phenomenon is likely to happen. As the 
upper limit of the excursions in $\dot{M}$ goes beyond the critical 
cross-over point from supersonic to subsonic propeller regime, the 
pulsed fraction will decrease because that part of $\dot{M}$ which 
is above this critical point will not contribute to the pulsed
flux, and this decrease will increase with increasing $\dot{M}$. 
This may be a natural explanation for the above observation of 
reduced pulsed fraction at higher luminosity. More quantitative
considerations will be given elsewhere.         

\subsection{Comparison with magnetar model}
\label{sec:magnetar}

In a recent paper, P08 have described a model
in which 1E is a \emph{magnetar}, \ie, a neutron star with a 
superstrong magnetic field $\sim 10^{15}$ G with a low-mass 
companion. The 6.7 hr period is interpreted in this model as the 
spin period of the neutron star, the idea being that a neutron 
star with such strong magnetic field as above can be spun down to
such long spin period, or such low spin frequency, in $\sim 2000$
yrs. Magnetars are a fascinating possibility, and their relevance
to soft gamma repeaters (SGRs) and possibly to anomalous X-ray 
pulsars (AXPs) has been the subject of much recent study. 
P08 have invoked an analogy with polars or AM Her-type 
cataclysmic variables containing white dwarfs with unusually 
strong magnetic fields, wherein torques acting on the magnetar
spin it down in a short time to spin periods in close
synchronism with the binary orbital period. In this analogy,
they have been inspired by the similarity of the shape the 1E
light curve to those of AM Her systems.   

We have desribed in this work a model which does not require a
neutron star with a superstrong magnetic field, but rather 
interprets the 6.7 hr period as the orbital period of the binary
system consisting of a neutron star with a canonical magnetic
field of $\sim 10^{12}$ G with a low-mass companion, the newborn,
fast-rotating neutron star being in the propeller phase, and the 
propeller emission being modulated in the eccentric orbit of a
young post-SN binary. We find that the observed 1E light curve 
can be quantitatively accounted for by our model. Our analogy 
is with propeller regimes of SXRTs like Aquila X-1 in their 
low/quiescent states, which we consider to be old, circularized 
analogues of 1E which are no longer orbitally modulated, but 
which have remarkably similar spectral properties. In this analogy,
we have been inspired by the similarity between 1E and the SXRTs in 
both the spectral characteristics and their changes with source 
luminosity, as well as the shapes of the outbursts and the way 
in which propeller-like properties emerge at low luminosities 
during outburst decays. 

An interesting question is that of possible discriminators 
between the above two models. It appears to us that if all 
observed properties of 1E and similar systems can be accounted 
for by known characteristics of early stages of pre-LMXBs born 
according to the standard CE evolution and He-star supernova 
scenario, such as we have described in this paper (or by other
possible models involving standard evolutionary scenarios),
there would not be any compelling need for invoking exotic objects 
like magnetars for this class of objects. On the other hand, if 
one finds unique observed features in this class of objects that 
cannot be explained at all within the framework of standard 
evolutionary scenarios, presence of magnetars in such objects may 
well be hinted at. However, answering this question is beyond the 
scope of this paper: we are pursuing the matter, and the results 
will be reported elsewhere.    

\subsection{Conclusions}
\label{sec:conclusion}

The work reported here suggests that 1E-type systems are early
stages of pre-LMXBs born in the SN of He-stars in binaries of
(He-star + low-mass star) produced by common-envelope
(CE) evolution. As long as the post-SN binary is eccentric, and
the neutron star is in the propeller regime, soft X-ray emission
modulated at the orbital period may be expected to occur. As 
the orbit circularizes, modulation would stop, and as the low-mass
companion moves out of Roche-lobe contact, the source would not
be observed in X-rays. The companion would come into Roche-lobe
contact again on a long timescale due to orbit shrinkage by
emission of gravitational radiation and magnetic braking, and/or
by the evolutionary expansion of the companion. This would lead
to a standard LMXB: an old neutron star in circular orbit with
a low-mass companion. Thus, steady-state arguments, with lifetimes
of 1E-type systems estimated at $\sim 10^6-10^7$ yrs and those
of LMXBs estimated at $\sim 10^8-10^9$ yrs, would lead us to expect
$\sim 1$ 1E-type systems per $\sim 100$ LMXBs, which is consistent
with current observations. However, we must be careful here, as 
these are overall arguments for the whole population. If one 
specifically investigates young supernova remnants (SNRs), the 
chances of finding such systems may be considerably higher, since
eccentric binary systems are to be found preferentially in such 
SNRs. More detailed considerations will be given elsewhere.

The lifetime of the eccentric-binary phase may be increased by an
effect we have not included in this introductory work. The effect
is that of an \emph{enhancement} of eccentricity when mass and
angular momentum are lost from a binary system which is already
eccentric. This dynamical effect is well-known in the literature
(see, \eg, \cite{huang}) and its applications to compact X-ray 
binaries have been made earlier (\cite{ghoshdcir}). For an
eccentric compact binary with the neutron star in the propeller 
regime leading to the loss of both mass and angular momentum from 
the system, such considerations are applicable. However, it is 
possible that, at the rates of mass transfer and loss inferred 
for 1E-type systems, this effect is a minor one.       

Several lines of further investigation are naturally suggested by
the considerations we have given in this paper. Foremost among 
them is a theory of the spectral characteristics of propeller
emission in disk-fed propeller systems. This would help clarify 
the remarkable spectral similarity (including changes in spectral
parameters with luminosity) between 1E and SXRTs like Aquila X-1 
in their low/quiescent state, as described in 
Sec.~\ref{sec:spectra}. A search for point soft X-ray sources in 
other young SNRs would clarify the observational situation greatly. 
We note that these sources may or may not be periodically modulated, 
as we have argued in Sec.~\ref{sec:ldepltcurve} that such 
modulations may decrease and disappear in certain luminosity 
states. However, the spectral characteristics would still be a most 
valuable diagnostic. These and other investigations are
under way, and results will be reported elsewhere.

\begin{acknowledgements}
      It is a pleasure to thank A. de Luca for sending data 
on the light curves, to thank E. P. J. van den Heuvel and 
L. Stella for stimulating discussions, and to thank the referee 
for comments which improved the paper considerably.
\end{acknowledgements}


\begin{thebibliography}{}

\bibitem[Avni~1976]{avni} Avni Y.  1976, ApJ, 209, 574-577
\bibitem[Bath~\&~Pringle~1981]{bath} Bath G., Pringle J. 1981, MNRAS, 194, 967-986
\bibitem[Brown~\&~Boyle 1984]{BB} Brown J., Boyle C. 1984, Astron \&
  Astrophys, 141, 369-375
\bibitem[Campana~\etal~1998]{campana} Campana S. \etal 1998, ApJ, 499, L65-L68
\bibitem[Davies~\etal~1979]{dav1} Davies R. \etal 1979, MNRAS, 186, 779-782
\bibitem[1981]{dav2} Davies R., \etal, 1981, MNRAS, 196, 209-224
\bibitem[de~Luca~\etal~2006]{lucasc} de Luca A. \etal, 2006, Science,
  313, 814-817 (dL06)
\bibitem[de~Luca~\etal~2008]{lucair} de Luca A. \etal, 2008, ApJ, 682, 1185-1194
\bibitem[Faulkner~1971]{faulkner} Faulkner J. 1971, ApJ, 170, L99-L104
\bibitem[Ghosh~\etal~1981]{ghoshdcir} Ghosh P. \etal 1981, ApJ, 251, 230-245.
\bibitem[Ghosh~1995]{ghoshprop} Ghosh P. 1995, ApJ, 453, 411-418 (G95)
\bibitem[Ghosh~2007]{ghoshbook} Ghosh P. 2007, ``Rotation and Accretion
  Powered Pulsars'' (Singapore: World Scientific Publications) 
\bibitem[Heyl~\&~Hernquist~2002]{heyl} Heyl J., Hernquist H. 2002,
  ApJ, 567, 510-514
\bibitem[Huang~1963]{huang} Huang S. -S. 1963, ApJ, 138, 471
\bibitem[Illarionov~\&~Kompaneets 1990]{ik} Illarionov A. F., Kompaneets D. A. 1990, 
MNRAS, 247, 219
\bibitem[Illarionov~\&~Sunyaev 1975]{is} Illarionov A. F., Sunyaev R. A. 1975, 
Astron \& Astrophysics, 39, 185-196 (IS75)
\bibitem[Illarionov~\etal~1993]{iik} Illarionov A. F.,
  Igumenschev I. V., Kompaneets D. A. 1993, 'The evolution of X-ray
  Binaries', ed. S. S. Holt \& C. S. Day (new York: AIP), 601
\bibitem[Joss~\&~Rappaport~1984]{joss} Joss P., Rappaport S. 1984, Ann. Rev. Astron 
Astrophys, 22, 537-592
\bibitem[Kalogera~1996]{vicky} Kalogera V. 1996, ApJ, 471, 352-365
\bibitem[Landin \etal~2009]{landin} Landin N. R., Mendez L. T. S.,
  Vaz L. P. R., Astron \& Astrophys, 494, 209-227. 
\bibitem[Lightman~1974]{lightman} Lightman A. 1974, ApJ, 194, 429-437.
\bibitem[ Lovelace \etal~1999]{love} Lovelace, R. V. E., Romanova,
  M. M., Bisnovatyi-Kogan, G. S. 1999, ApJ, 514, 368-372 (LRB99)
\bibitem[Lubow~\&~Shu~1975]{lubow} Lubow S.,  Shu F. 1975, ApJ, 198, 383-405
\bibitem[Lynden-Bell~\&~Pringle~1974]{lynden} Lynden-Bell D., 
  Pringle F. 1974, MNRAS, 168, 603-637
\bibitem[Meibom~\&~Mathieu 2005]{MM} Meibom S., Mathieu R. D. 2005, ApJ, 620, 970-983.
\bibitem[Mignani~\etal~2007]{mignani} Mignani R., et. al. 2007. ArXiv
  Astro-ph : 0711.0113
\bibitem[Mineshige~\etal~1991]{mrf} Mineshige S., Rees M. J.,
  Fabian A. C. 1991, MNRAS, 251, 555
\bibitem[Pavlov~\etal~2002]{pavlov} Pavlov G. \etal, 2002, 'Neutron Stars in Supernova
  Remnants' ASP Conference Series. Vol 271, 2002.
\bibitem[Pizzolato~\etal~2008]{pizzo} Pizzolato F. \etal, 2008, ApJ,
  681, 530-542 (P08)
\bibitem[Pravdo~\&~Ghosh~2001]{ghosh} Pravdo S., Ghosh P. 2001, ApJ, 554, 383-390
\bibitem[Pringle~1981]{pringle} Pringle J. 1981,
  Ann. Rev. Astron. Astrophys. 19, 137-62
\bibitem[Romanova \etal~2004]{rom04} Romanova, M. M., Ustyugova,
  G. V., Koldoba, A. V., Lovelace, R. V. E. 2004, ApJ, 616, L151-L155 
\bibitem[Romanova \etal~2005]{rom05} Romanova, M. M., Ustyugova,
  G. V., Koldoba, A. V., Lovelace, R. V. E. 2005, ApJ, 635, L165-L168 (RUKL05)
\bibitem[Shakura~\&~Sunyaev~1973]{shakura} Shakura N., Sunyaev
  R. 1973, Astron \& Astrophys, 24, 337-355
\bibitem[Stella~\etal~2000]{stella}  Stella L. \etal 2000, ApJ, 537, L115-L118
\bibitem[Touhy~\&~Garmire~1980]{garmire} Tuohy I., Garmire G. 1980,
  ApJ, 239, L107-L110
\bibitem[Ustyugova \etal~2006]{usty} Ustyugova, G. V.,
  Koldoba, A. V., Romanova, M. M., Lovelace, R. V. E. 2006, ApJ, 646,
  304-318 (UKRL06)
\bibitem[Verbunt~\&~Zwaan~1981]{verbunt} Verbunt F., Zwaan C. 1981, 
Astron \& Astrophys, 100, L7-L9
\bibitem[Zahn~1977]{zahn}  Zahn J.-P. 1977, Astron \& Astrophys, 57,
  383-394
\bibitem[Zahn~1978]{zahne}  Zahn J.-P. 1978, Astron \& Astrophys, 67,
  162
 
\end{thebibliography}
\end{document}